\documentclass[%
 aip,
 amsmath,amssymb,
 reprint,%
]{revtex4-1}

\usepackage{pdfpages}
\makeatletter
\AtBeginDocument{\let\LS@rot\@undefined}
\makeatother

\usepackage{tabularx}
\usepackage{dcolumn}
\usepackage{xcolor}
\usepackage{graphicx}
\usepackage{geometry}
\usepackage{gensymb}
\usepackage{cleveref}
\usepackage{bm}
\usepackage[version=4]{mhchem}
\usepackage{multirow}
\usepackage[utf8]{inputenc}
\usepackage[T1]{fontenc}
\usepackage{mathptmx}
\usepackage{footnote}
\usepackage{url}

\graphicspath{ {./fig/} }

\usepackage{geometry}
\begin{document}

\preprint{AIP/123-QED}

\title{Differentiable thermodynamic modeling}

\author{Pin-Wen Guan}
\email{pinweng@andrew.cmu.edu}
\affiliation{Department of Mechanical Engineering, Carnegie Mellon University, Pittsburgh, Pennsylvania 15213, USA}

\date{\today}

\begin{abstract}

A new framework of thermodynamic modeling is proposed by introducing the concept of differentiable programming, where all the thermodynamic observables including both thermochemical quantities and phase equilibria can be differentiated with respect to the underlying model parameters, thus allowing the models learned by gradient-based optimization. It is shown that thermodynamic modeling and deep learning can be seamlessly integrated and unified within this framework. A preliminary successful application is demonstrated for the Cu-Rh system. It is expected that thermodynamic modeling in a deep learning style can increase prediction power of models, and provide more effective guidance for design, synthesis and optimization of multi-component materials with complex chemistry via learning various types of data.
\end{abstract}

\maketitle

Various phenomena in materials have their roots in thermodynamics, of which the fundamental laws have been well established very early. However, thermodynamics of materials is far from being well studied, due to the vast number of degrees of freedom including composition, temperature, pressure, strain, external field and dimensionality (e.g., bulk vs nano). In addition, the underlying microscopic mechanisms contributing to macroscopic thermodynamic observables are diverse, including lattice disorder, atomic vibration, electronic excitation, magnetic excitation, etc., making accurate and comprehensive theoretical descriptions difficult. Therefore, developing a sufficiently efficient methodology is highly crucial for thermodynamic modeling to guide design and synthesis of materials more effectively. For this end, there have emerged many encouraging progresses in recent years. One notable direction is the usage of machine learning (ML) techniques \cite{schmidt2017predicting, ye2018deep, jha2018elemnet, ubaru2017formation, teichert2019machine, lapointe2020machine, ryan2018crystal, kaufmann2020discovery, zhang2020phase, Huang2019, pilania2015structure, seko2014machine, guan2020meltnet}, the generalizability of which allows predictions based on limited amount of data. So far, the methods along this direction can be classified into two types based on the types of training data and predicted quantities of the ML models. The type-I models are trained on and predict thermochemical quantities\cite{schmidt2017predicting, ye2018deep, jha2018elemnet, ubaru2017formation, teichert2019machine, lapointe2020machine, ryan2018crystal, kaufmann2020discovery}, while the type-II models are trained on and predict phase equilibria \cite{zhang2020phase, Huang2019, pilania2015structure, seko2014machine, guan2020meltnet}. Naturally, one may ask if there can be a cross-type learning, i.e., learning thermochemical quantities from phase equilibria (or learning phase equilibria from thermochemical quantities, which is essentially within type-I, since the former can be derived if the latter are fully determined, thus this type is ignored here). Such type of learning has the following advantages. First, it is quite often that thermochemical data are scarce or not reliable enough, thus training on phase equilibria is the desirable or even the only option. Second, prediction of thermochemical quantities like Gibbs energies allows calculations of not only phase equilibria, but also useful parameters such as thermodynamic factor in diffusion. Third, phase equilibria calculated based on thermochemical quantities are more physics-based compared with direct predictions by ML. Therefore, it is of great practical and scientific interests to develop this new type of ML paradigm for thermodynamics.

In this work, a new framework capable of learning thermodynamic potentials from both thermochemical data and phase equilibrium data is proposed, and it will be shown that thermodynamic modeling and deep learning are seamlessly unified within this framework. This is achieved by introducing differentiable programming (DP) \cite{baydin2018automatic}, a programming paradigm where the computation flow can be differentiated throughout via automatic differentiation, thus allowing gradient-based optimization of parameters in the flow. Due to its ability to incorporate physics into deep learning, DP has increasing applications in different fields recently, including molecular dynamics \cite{schoenholz2020jax}, tensor networks \cite{liao2019differentiable}, quantum chemistry \cite{tamayo2018automatic} and density functional theory (DFT) \cite{li2021kohn}.

Consider a set of phases $\{G_{\theta_i}\}$. The thermodynamic potential of the phase $\theta_i$ can be represented as
\[G_{\theta_i}(F(x),C;A_{\theta_i})\]
which is a function of the descriptor $F(x)$ based on composition $x$, external conditions $C$ (e.g., temperature and pressure) with parameters $A_{\theta_i}$. Usage of $F(x)$ is also called feature engineering. When $F$ is identity, i.e., $F(x)=x$, raw composition is directly used. $G_{\theta_i}$ can be in various forms that are differentiable, such as the conventional one based on polynomials \cite{redlich1948algebraic, Lukas2007} and the more ML-oriented one based on deep neural networks \cite{teichert2019machine}.

Once $\{G_{\theta_i}\}$ of all the relevant phases are known, each thermochemical quantity $g_j$ and phase equilibrium $e_j$ can then be calculated, i.e.,
\[\widehat{g_j}=\widehat{g_j}(\{G_{\theta_i}\}), \widehat{e_j}=\widehat{e_j}(\{G_{\theta_i}\})\]
which are functionals of $\{G_{\theta_i}\}$. The loss function can then be computed as
\[L=\sum_{j}l(g_j,\widehat{g_j})+\sum_{j}l(e_j,\widehat{e_j})\]
where $l$ is a function measuring the difference between two values, with a common choice to be the squared error $l(g_j,\widehat{g_j})=(g_j-\widehat{g_j})^2$. Finally, the parameter set $\{A_{\theta_i}\}$ of thermodynamic potentials can be obtained by minimizing the loss function
\[\{A_{\theta_i}\}=arg\min_{\{A_{\theta_i}\}}L\]
The minimization of the loss function $L$ relies on calculating its gradient $\nabla_{\{A_{\theta_i}\}}L$, which is made possible by DP. Obviously, the most non-trivial part in the above computation flow for differentiation is the phase equilibrium calculation for $e_j$, which generally requires minimization of the thermodynamic potential of the whole system, the major subject of research in computational thermodynamics. The minimization can be usually divided into two main steps \cite{piro2016global}. The first step is a global minimization where the thermodynamic potential surface is sampled and an initial solution is generated. The second step is refining calculations to obtain the final solution satisfying all the equilibrium conditions. 

\begin{figure*}
\includegraphics[width=\textwidth]{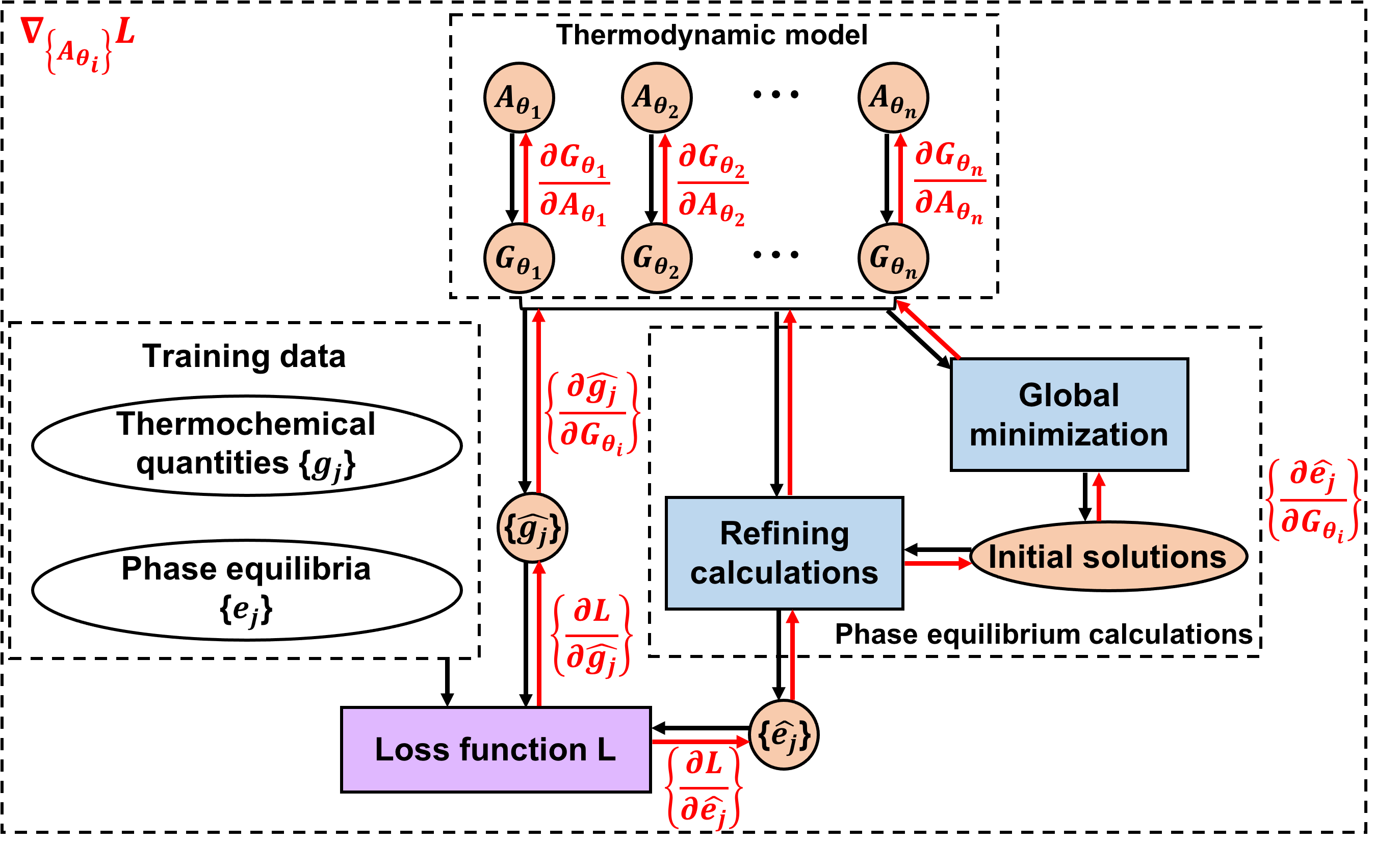}
\centering
\caption{Scheme of differentiable thermodynamic modeling. The forward computation flow is indicated by black arrows. The backpropagation gradient flow is indicated by red arrows, which is used to calculate the gradient of the loss function with respect to model parameters, $\nabla_{\{A_{\theta_i}\}}L$, for minimizing the loss function. The variable parts in the differentiation are filled with colors, while the invariable parts (the training data) have no filled colors.} 
\label{fig:flow}
\end{figure*}

The scheme of differentiable thermodynamic modeling is shown in \cref{fig:flow}. The forward mode is essentially similar to a routine thermodynamic calculation. In the backward mode, the computations of gradients propagate from the loss function towards the model parameters through a series of intermediate steps, and all the single steps are finally assembled together to obtain $\nabla_{\{A_{\theta_i}\}}L$ based on elementary rules of differentiation. It can be seen that such scheme is very similar to that in deep learning. In fact, differentiable programming can be regarded an extension of deep learning with the use of diverse differentiable components beyond classical neural networks.

The above general framework has been implemented for a simple two-phase binary system Cu-Rh for demonstration, which has a liquidus and a solidus between the fcc phase and the liquid (liq) phase besides a fcc miscibility gap right below the solidus. To calculate phase equilibria under given temperature and pressure, Gibbs energy is used as the thermodynamic potential. The four-parameter model by Lesnik et al. \cite{chakrabarti1982cu} is taken as the target to learn:
\begin{multline*}
G_{\theta}=x_{\theta}{^{\circ}G_{\theta,Rh}}+(1-x_{\theta}){^{\circ}G_{\theta,Cu}}\\
+RT[xlnx+(1-x)ln(1-x)]+x(1-x)(A_{\theta}+A_{\theta}^{'}x)\\
\end{multline*}
where $\theta$=fcc,liq, $R$ is the gas constant, $x_{\theta}$ is the Rh composition in the $\theta$ phase, and $^{\circ}G_{\theta,Rh}$ and $^{\circ}G_{\theta,Cu}$  are the standard Gibbs energies of Rh and Cu in the $\theta$ phase, respectively. The four model parameters have been assessed as $(A_{fcc},A_{fcc}^{'},A_{liq},A_{liq}^{'})=(14.609,11.051,8.414,19.799)$ (in kJ/mol), which are regarded as the true values in the present work. From this set of model parameters, the training data is generated, including phase boundaries at 1000-2200 K.

The Gibbs energy minimization is divided into two steps. The first step is a global minimization on a grid of compositions based on a convex-hull algorithm, generating an approximate solution which is to be refined in the second step. The second step is an iterative self-consistent procedure, where the Newton–Raphson method is used to solve phase equilibria under fixed chemical potential which is then updated based on the newest solved phase equilibria in each iteration, and the iterations stop when convergence is achieved.  The loss function for this example system is defined as
\[L=\sum_{i}L_{liq-fcc,i}+\sum_{i}L_{fcc-fcc,i}\]
with
\begin{multline*}
  L_{liq-fcc,i}=\\
    \begin{cases}
      \alpha[(\widehat{x_{liq}}(T_i)-x_{liq}(T_i))^2+(\widehat{x_{fcc}}(T_i)-x_{fcc}(T_i))^2]\\   \hfill  \text{if $\widehat{x_{liq}}(T_i)$ and $\widehat{x_{fcc}}(T_i)$ computable}
      \\
      (\min\limits_{x}\frac{DF}{RT_i})^2 \hfill \text{otherwise}
    \end{cases}      
\end{multline*}
and
\begin{multline*}
  L_{fcc-fcc,i}=\\
    \begin{cases}
      \alpha[(\widehat{x_{fcc\#1}}(T_i)-x_{fcc\#1}(T_i))^2+(\widehat{x_{fcc\#2}}(T_i)-x_{fcc\#2}(T_i))^2]\\   \hfill  \text{if $\widehat{x_{fcc\#1}}(T_i)$ and $\widehat{x_{fcc\#2}}(T_i)$ computable}
      \\
      [\mathrm{ReLU}(\frac{1}{RT_i}\min\limits_x\frac{\partial^2G_{fcc}(x,T_i)}{\partial x^2})]^2 \hfill \text{otherwise}
    \end{cases}      
\end{multline*}
where the rectified linear unit ReLU$(z)=\max(0,z)$, and $\alpha$ is a scaling factor for improving convergence in minimization of the loss function. In the present case, $\alpha=100$ is used. The driving force DF of the metastable phase is the distance in terms of Gibbs energy between the stable tangent plane and a tangent plane parallel to the metastable phase \cite{Lukas2007}. The above loss function contains two contributions from the liquid-fcc equilibrium and the fcc miscibility gap respectively. Since the target type of phase equilibrium may not be correctly reproduced if the model parameters have large deviation from their target values, penalties are imposed instead in such scenarios to favor the regions where $\min\limits_{x}DF=0$ and $\min\limits_{x}\frac{\partial^2G_{fcc}(x,T_i)}{\partial x^2})]^2<0$, i.e., the liquid-fcc equilibrium and the fcc miscibility gap exist at some compositions. 

A differentiable program for calculating the above loss function is written using JAX \cite{schoenholz2020jax}, a machine learning library which can automatically differentiate Python and NumPy functions. Notably, JAX can differentiate through control flows like loops and branches, which are key structures in the Gibbs energy minimization. The gradient of the loss function, $\nabla_{\{A_{\theta_i}\}}L$, is then obtained by automatic differentiation of the program. Given its gradient, the loss function is minimized using a gradient-based optimization algorithm, the L-BFGS-B method.   

\begin{figure*}
\includegraphics[width=\textwidth]{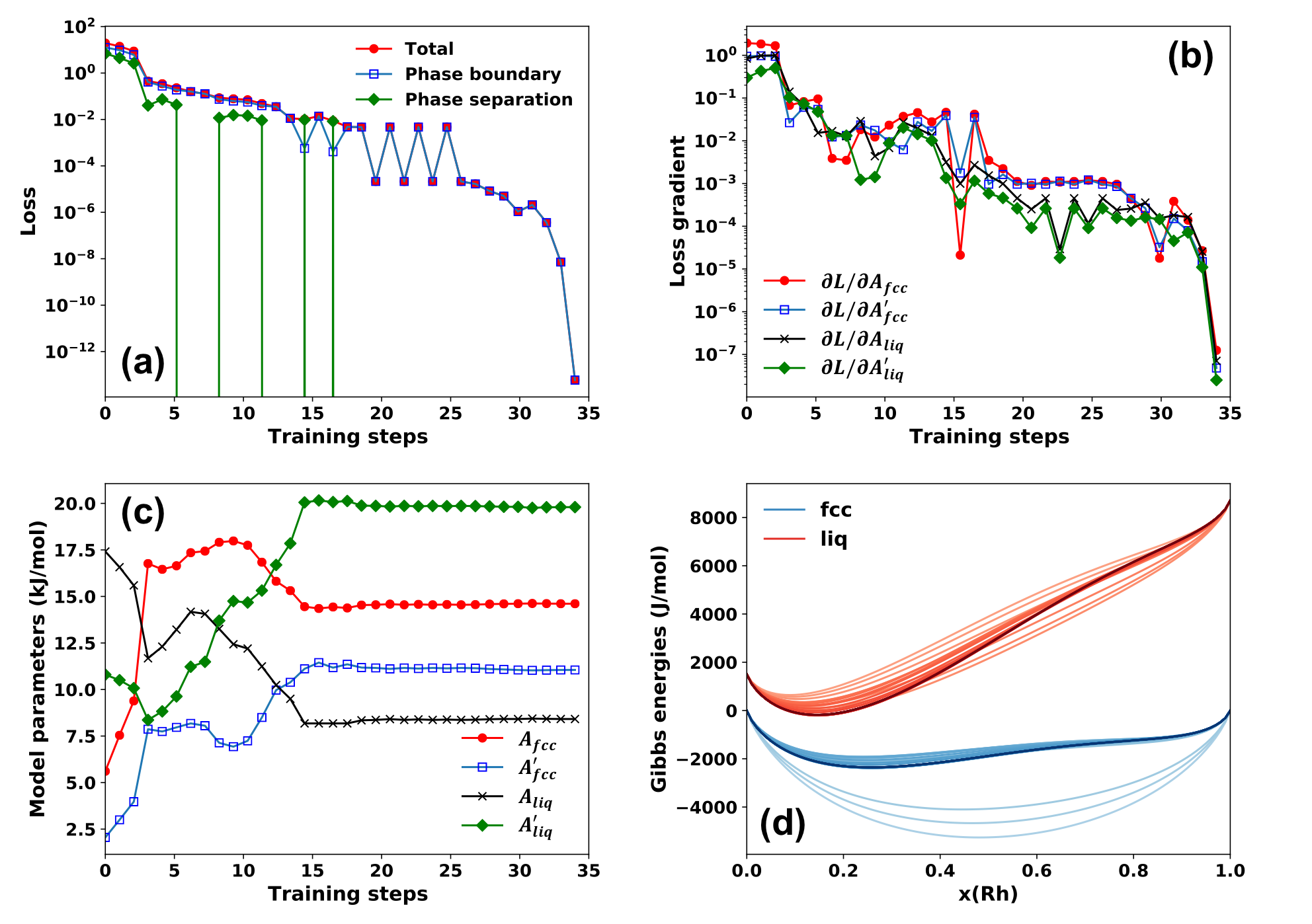}
\centering
\caption{Evolution of various quantities in the training process for the Cu-Rh model system: (a) loss function and the contributions from different origins. (b) gradient of loss function. (c) model parameters. (d) Gibbs energies of involved phases at 1200 K. Heavier colors are corresponded to larger training steps.} 
\label{fig:train}
\end{figure*}

\begin{figure}
\includegraphics[width=\linewidth]{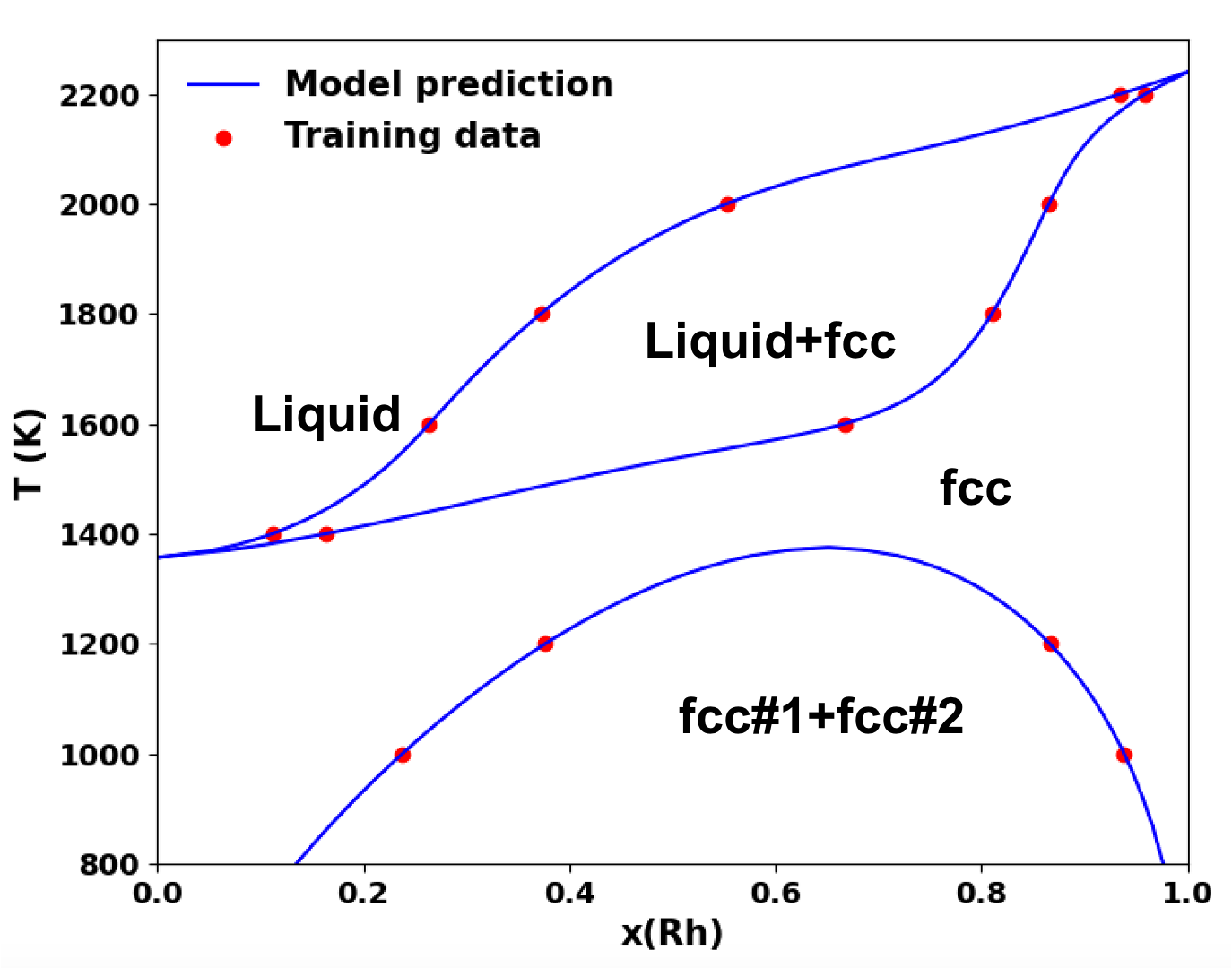}
\centering
\caption{Phase diagram of the Cu-Rh system predicted by the trained model, in comparison with the training data.} 
\label{fig:pd}
\end{figure}

The training process for the Cu-Rh model system is shown in \cref{fig:train}. Despite starting with quite unreasonable initial model parameters, the minimization of the loss function is quite efficient, with good convergence achieved within a few tens of steps, which is made possible by the explicitly calculated gradient of the loss function. In this case, the liquid-fcc equilibrium always exists during training, thus the loss function has only two non-zero contributions from displacement of the phase boundaries (liquid-fcc and fcc-fcc) and absence of the phase separation (fcc miscibility gap) caused by inaccurate model parameters, respectively. The latter contribution vanishes after the fcc miscibility gap is made exist, and then the training is only accompanied with quantitative adjustment of the phase boundaries. With the loss function minimized, its four-component gradient driving the training process approaches the zero vector, and the model parameters also converges to the true values. To better understand how the thermodynamic model evolves, the trajectories of Gibbs energies of the two involved phases at 1200 K in the training are plotted. In consistence with initial absence of the fcc miscibility gap, the Gibbs energy of the fcc phase is initially a convex function without spinodal decomposition, but gradually trained to be non-convex leading to phase separation. The phase diagram of the Cu-Rh system predicted by the trained model is shown in \cref{fig:pd}, along with the training data. It can be seen that the predicted phase diagram is in excellent agreement with the training data, meaning the present model training is highly successful.

The above example provides a preliminary successful application of differentiable thermodynamic modeling, which is able to learn thermodynamics from mixed types of data on thermochemistry and phase equilibria in a deep learning style. Due to simplicity of the binary system and limited amount of training data, a polynomial with raw elemental compositions directly used as inputs is a suitable form to represent the excess Gibbs energy of each phase, which is also the routine practice in conventional thermodynamic modeling. However, such representation may suffer “curse of dimensionality” in high-dimension chemical space. For instance, considering an extreme case where a phase contains 100 elements, there are 100 compositional variables, $\mathrm{C}_{100}^2=4950$ binary interactions, $\mathrm{C}_{100}^3$=161700 ternary interactions and $\mathrm{C}_{100}^4$=3921225 quaternary interaction, totaling a daunting number of model parameters with each interaction represented by a conventional parameterized polynomial. It is therefore desirable to explore an alternative representation of the Gibbs energy. Due to its strong expressivity, the neural network has emerged as a promising tool for this purpose, and there have been quite a few related works in the literature. Using elemental compositions as inputs, a deep neural network trained on DFT data has achieved a mean average error of 0.05 eV/atom in predicting formation enthalpies \cite{jha2018elemnet}. There have been also efforts in introducing physical attributes such like electronegativity and atomic radius into inputs by feature engineering to alleviate the difficulty due to vast dimensions of the composition space \cite{ward2016general}. This is still a field under active research, and further discussions are beyond the scope of the present work. However, it is quite obvious that the present framework of differentiable thermodynamic modeling can provide a necessary platform for introducing neural networks to learn thermodynamics from diverse types of data. 

From the perspective of mapping, a set of phase equilibrium data is actually a sample from the map $f:(x,C)\rightarrow e$, where $e$ is the phase equilibrium at composition $x$ and external condition $C$. To incorporate more physics, thermodynamic potentials $\{G_{\theta}\}$ is introduced as intermediate variables with two maps $f_1:(x,C)\rightarrow \{G_{\theta}\}$ and $f_2:\{G_{\theta}\}\rightarrow e$, which are usually called “thermodynamic model” and “phase equilibrium calculation”, respectively. The map $f$ is just their composition:
\[f=f_2 \circ f_1\]
Note that $f_1$ and $f_2$ are very different in nature. $f_1$ is usually complicated and sometimes even obscure, packaging the whole physics of each single-phase material that is difficult to calculate explicitly without capturing all the atomic and electronic details, but this is just the part deep learning can find its largest use. In contrast, $f_2$ is more straightforward, thus most suitable for a direct physical computation. Differentiable thermodynamic modeling offers a seamless integration of these two components. 

In summary, the present work proposes a deep learning framework for thermodynamic modeling, which is termed differentiable thermodynamic modeling. It is based on differentiable programming, which allows differentiation throughout the computation flow and therefore gradient-based optimization of parameters. Under this framework, thermodynamics can be learned from different types of data on thermochemistry and phase equilibria in a deep learning style, and thermodynamic modeling and deep learning are de facto unified and indistinguishable. Its preliminary success is demonstrated by application in training a model for the Cu-Rh system. It is expected that differentiable thermodynamic modeling can facilitate exploration of thermodynamics of multi-component systems with complex chemistry, as well as design, synthesis and optimization of multi-component materials. 

\section*{Research data}

The data that support the findings of this study are available from the corresponding author upon reasonable request. The data and code used in this work will be also made publicly available on Github.

\section*{Acknowledgement}

The author would like to acknowledge interesting presentations and discussions from the Scientific Machine Learning Webinar Series (https://www.cmu.edu/aced/sciML.html). Acknowledgment is also made to the Extreme Science and Engineering Discovery Environment (XSEDE) for providing computational resources through Award No. TG-CTS180061.

\clearpage

\bibliography{main}

\end{document}